\documentclass[aps,prltightenlines,superscriptaddress,nofootinbib,amsmath,amssymb,reprint]{revtex4-2}
 
\usepackage{lineno} 
\usepackage{graphicx}
\usepackage{dcolumn}
\usepackage{bm}
 \usepackage{verbatim}

\RequirePackage{color}

\begin{document}

\preprint{AIP/123-QED}

\title[Sample title]{Evidence of Coherent Elastic Neutrino-Nucleus Scattering with COHERENT’s Germanium Array}

\newcommand{\WJCdesc}{\affiliation{Washington \& Jefferson College, Washington, PA, 15301, USA}}
\newcommand{\SNUdesc}{\affiliation{Department of Physics and Astronomy, Seoul National University, Seoul, 08826, Korea}}
\newcommand{\Dukedesc}{\affiliation{Department of Physics, Duke University, Durham, NC, 27708, USA}}
\newcommand{\TUNLdesc}{\affiliation{Triangle Universities Nuclear Laboratory, Durham, NC, 27708, USA}}
\newcommand{\Mephidesc}{\affiliation{National Research Nuclear University MEPhI (Moscow Engineering Physics Institute), Moscow, 115409, Russian Federation}}
\newcommand{\ITEPnewadesc}{\affiliation{National Research Center  ``Kurchatov Institute'' , Moscow, 123182, Russian Federation }}
\newcommand{\UTKdesc}{\affiliation{Department of Physics and Astronomy, University of Tennessee, Knoxville, TN, 37996, USA}}
\newcommand{\USDdesc}{\affiliation{Department of Physics, University of South Dakota, Vermillion, SD, 57069, USA}}
\newcommand{\NCSUdesc}{\affiliation{Department of Physics, North Carolina State University, Raleigh, NC, 27695, USA}}
\newcommand{\Sandiadesc}{\affiliation{Sandia National Laboratories, Livermore, CA, 94550, USA}}
\newcommand{\Tuftsdesc}{\affiliation{Department of Physics and Astronomy, Tufts University, Medford, MA, 02155, USA}}
\newcommand{\ORNLdesc}{\affiliation{Oak Ridge National Laboratory, Oak Ridge, TN, 37831, USA}}
\newcommand{\UWdesc}{\affiliation{Center for Experimental Nuclear Physics and Astrophysics \& Department of Physics, University of Washington, Seattle, WA, 98195, USA}}
\newcommand{\LANLdesc}{\affiliation{Los Alamos National Laboratory, Los Alamos, NM, 87545, USA}}
\newcommand{\CNLdesc}{\affiliation{Canadian Nuclear Laboratories Ltd, Chalk River, Ontario, K0J 1J0, Canada}}
\newcommand{\IUdesc}{\affiliation{Department of Physics, Indiana University, Bloomington, IN, 47405, USA}}
\newcommand{\VTdesc}{\affiliation{Center for Neutrino Physics, Virginia Tech, Blacksburg, VA, 24061, USA}}
\newcommand{\CMUdesc}{\affiliation{Department of Physics, Carnegie Mellon University, Pittsburgh, PA, 15213, USA}}
\newcommand{\NCCUdesc}{\affiliation{Department of Mathematics and Physics, North Carolina Central University, Durham, NC, 27707, USA}}
\newcommand{\NCSUnucengdesc}{\affiliation{Department of Nuclear Engineering, North Carolina State University, Raleigh, NC, 27695, USA}}
\newcommand{\FSUdesc}{\affiliation{Department of Physics, Florida State University, Tallahassee, FL, 32306, USA}}
\newcommand{\UFdesc}{\affiliation{Department of Physics, University of Florida, Gainesville, FL, 32611, USA}}
\newcommand{\Concorddesc}{\affiliation{Department of Physical and Environmental Sciences, Concord University, Athens, WV, 24712, USA}}
\newcommand{\SLACdesc}{\affiliation{SLAC National Accelerator Laboratory, Menlo Park, CA, 94025, USA}}
\newcommand{\Laurentiandesc}{\affiliation{Department of Physics, Laurentian University, Sudbury, Ontario, P3E 2C6, Canada}}
\author{S.~Adamski}\WJCdesc
\author{M.~Ahn}\SNUdesc
\author{P.S.~Barbeau}\Dukedesc\TUNLdesc
\author{V.~Belov}\Mephidesc\ITEPnewadesc
\author{I.~Bernardi}\UTKdesc
\author{C.~Bock}\USDdesc
\author{A.~Bolozdynya}\Mephidesc
\author{R.~Bouabid}\Dukedesc\TUNLdesc
\author{J.~Browning}\NCSUdesc
\author{B.~Cabrera-Palmer}\Sandiadesc
\author{N.~Cedarblade-Jones}\Dukedesc\TUNLdesc
\author{A.I.~Col\'on Rivera}\Dukedesc\TUNLdesc
\author{E.~Conley}\Dukedesc
\author{V.~da Silva}\Tuftsdesc
\author{J.~Daughhetee}\ORNLdesc
\author{J.~Detwiler}\UWdesc
\author{K.~Ding}\USDdesc
\author{M.R.~Durand}\UWdesc
\author{Y.~Efremenko}\UTKdesc\ORNLdesc
\author{S.R.~Elliott}\LANLdesc
\author{A.~Erlandson}\CNLdesc
\author{L.~Fabris}\ORNLdesc
\author{A.~Galindo-Uribarri}\ORNLdesc\UTKdesc
\author{M.P.~Green}\TUNLdesc\ORNLdesc\NCSUdesc
\author{J.~Hakenm\"uller}\email{janina.hakenmuller@duke.edu}\Dukedesc
\author{M.R.~Heath}\ORNLdesc
\author{S.~Hedges}\altaffiliation{Also at: Lawrence Livermore National Laboratory, Livermore, CA, 94550, USA}\Dukedesc\TUNLdesc
\author{H.~Jeong}\SNUdesc
\author{B.A.~Johnson}\IUdesc
\author{T.~Johnson}\Dukedesc\TUNLdesc
\author{H.~Jones}\NCSUdesc
\author{A.~Khromov}\Mephidesc
\author{D.~Kim}\SNUdesc
\author{A.~Konovalov}\altaffiliation{Also at: Lebedev Physical Institute of the Russian Academy of Sciences, Moscow, 119991, Russian Federation}\Mephidesc
\author{E.~Kozlova}\Mephidesc
\author{A.~Kumpan}\Mephidesc
\author{O.~Kyzylova}\VTdesc
\author{Y.~Lee}\SNUdesc
\author{G.~Li}\CMUdesc
\author{L.~Li}\Dukedesc\TUNLdesc
\author{J.M.~Link}\VTdesc
\author{J.~Liu}\USDdesc
\author{M.~Luxnat}\IUdesc
\author{A.~Major}\Dukedesc
\author{K.~Mann}\NCSUdesc
\author{D.M.~Markoff}\NCCUdesc\TUNLdesc
\author{J.~Mattingly}\NCSUnucengdesc
\author{J.~Moye}\CMUdesc
\author{P.E.~Mueller}\ORNLdesc
\author{J.~Newby}\ORNLdesc
\author{N.~Ogoi}\NCCUdesc\TUNLdesc
\author{J.~O'Reilly}\Dukedesc
\author{D.S.~Parno}\CMUdesc
\author{D.~P\'erez-Loureiro}\CNLdesc
\author{D.~Pershey}\FSUdesc
\author{C.G.~Prior}\Dukedesc\TUNLdesc
\author{J.~Queen}\Dukedesc
\author{R.~Rapp}\WJCdesc
\author{H.~Ray}\UFdesc
\author{O.~Razuvaeva}\Mephidesc\ITEPnewadesc
\author{D.~Reyna}\Sandiadesc
\author{G.C.~Rich}\TUNLdesc
\author{D.~Rudik}\altaffiliation{Also at: University of Naples Federico II, Naples, 80138, Italy}\Mephidesc
\author{J.~Runge}\Dukedesc\TUNLdesc
\author{D.J.~Salvat}\IUdesc
\author{J.~Sander}\USDdesc
\author{K.~Scholberg}\Dukedesc
\author{A.~Shakirov}\Mephidesc
\author{G.~Simakov}\Mephidesc\ITEPnewadesc
\author{W.M.~Snow}\IUdesc
\author{V.~Sosnovtsev}\Mephidesc
\author{M.~Stringer}\CNLdesc
\author{T.~Subedi}\Concorddesc
\author{B.~Suh}\IUdesc
\author{B.~Sur}\CNLdesc
\author{R.~Tayloe}\IUdesc
\author{K.~Tellez-Giron-Flores}\VTdesc
\author{Y.-T.~Tsai}\SLACdesc
\author{E.E.~van Nieuwenhuizen}\Dukedesc\TUNLdesc
\author{C.J.~Virtue}\Laurentiandesc
\author{G.~Visser}\IUdesc
\author{K.~Walkup}\VTdesc
\author{E.M.~Ward}\UTKdesc
\author{T.~Wongjirad}\Tuftsdesc
\author{Y.~Yang}\USDdesc
\author{J.~Yoo}\SNUdesc
\author{C.-H.~Yu}\ORNLdesc
\author{A.~Zaalishvili}\Dukedesc\TUNLdesc

\date{\today}

\begin{abstract}
We report the first detection of coherent elastic neutrino-nucleus scattering (CEvNS) on natural germanium, measured at the Spallation Neutron Source at Oak Ridge National Laboratory. The Ge-Mini detector of the COHERENT collaboration employs large-mass, low-noise, high-purity germanium spectrometers, enabling excellent energy resolution, and an analysis threshold of 1.5\,keV electron-equivalent ionization energy. We observe a on-beam excess of 20.6$^{+7.1}_{-6.3}$ counts with a total exposure of 10.22\,GWhkg and we reject the no-CEvNS hypothesis with 3.9\,$\sigma$ significance. The result agrees with the predicted standard model of particle physics signal rate within 2 $\sigma$.
\end{abstract}

\maketitle
\paragraph*{Coherent elastic neutrino-nucleus scattering on germanium.-}

The neutrino interaction with the largest cross section at solar and supernova neutrino energies, coherent elastic neutrino-nucleus scattering (CEvNS) \cite{Freedman:1973yd,akimov2017observation}, has only been measured on two nuclear targets to date \cite{akimov2017observation, akimov2022measurement, akimov2021first}.  
In CEvNS interactions, neutrinos with energies below $\sim$50\,MeV impart a low enough momentum transfer to a nucleus such that the wavelengths of scattering waves are larger than the size of the nucleus, allowing a coherent summation of in-phase scatters from individual nucleons and thus an enhancement of the cross section.

CEvNS was first proposed in 1974 \cite{Freedman:1973yd,kopeliovich1974isotopic}, but the keV-scale recoil energies of elastically scattered nuclei make detection challenging. The COHERENT Collaboration 
leverages the capabilities of the Spallation Neutron Source (SNS) at Oak Ridge National Laboratory (ORNL) as a stopped-pion neutrino source for measurements of CEvNS.
A 60\,Hz pulsed proton beam bombards a liquid mercury target, producing neutrons and large quantities of pions. 
The $\pi^+$ particles stop predominantly ($>$99\%) in the target and decay at rest, yielding a prompt population of $\nu_{\mu}$ neutrinos (emitted with the same time structure of the SNS proton pulse with $\sim$400\,ns full-width-at-half-maximum (FWHM)) and $\mu^+$ particles. The subsequent decay of each $\mu^+$ yields an additional delayed $\nu_e$ and $\bar{\nu}_{\mu}$, extending to $\sim8\,\mu$s due to the 2.2$\mu$s $\mu^+$ lifetime.
Stopped-pion sources provide neutrino energies of up to $\sim$53\,MeV, yielding larger fractions of recoils above detector energy thresholds than for other low-energy neutrino sources.
Additionally, the pulsed nature of the beam allows for a strong suppression of steady-state backgrounds.
The COHERENT collaboration has made the only unambiguous detections of CEvNS to date, on CsI[Na] inorganic and liquid argon scintillator detectors \cite{akimov2017observation,akimov2021first,akimov2022measurement}. All COHERENT detectors are located in Neutrino Alley, an SNS basement corridor with significantly diminished neutron flux compared to elsewhere in the target building.

High-purity germanium detectors are highly suited to measure CEvNS due to their excellent energy resolution and keV-scale energy thresholds, as well as inherently low internal backgrounds \cite{luke1989low,barbeau2007large,TEXONO:2014eky,CoGeNT:2012sne,bonet2023full}. Development of the P-type Point-Contact (PPC) detector configuration enables sub-keV energy thresholds by reducing detector capacitance ($\sim1$\,pF) and thus electronic noise; the subsequent evolution of the Inverted Coaxial Point-Contact (ICPC) form factor has enabled the construction of detectors with masses in excess of 2\,kg retaining this excellent noise performance \cite{cooper2011novel,jany2021fabrication}.
Here we present the first conclusive measurement of CEvNS on natural germanium (20.5\% $^{70}$Ge, 27.4\% $^{72}$Ge, 7.8\% $^{73}$Ge, 36.5\% $^{74}$Ge and 7.8\% $^{76}$Ge \cite{Geabundance}) using Ge-Mini, an array of eight 2.2\,kg ICPC detectors with low background cryostats manufactured by Mirion Technologies, deployed at the SNS and operated during the SNS neutron production period 
from June - August 2023.  

\begin{figure}[h]
\includegraphics[width=0.48\textwidth]{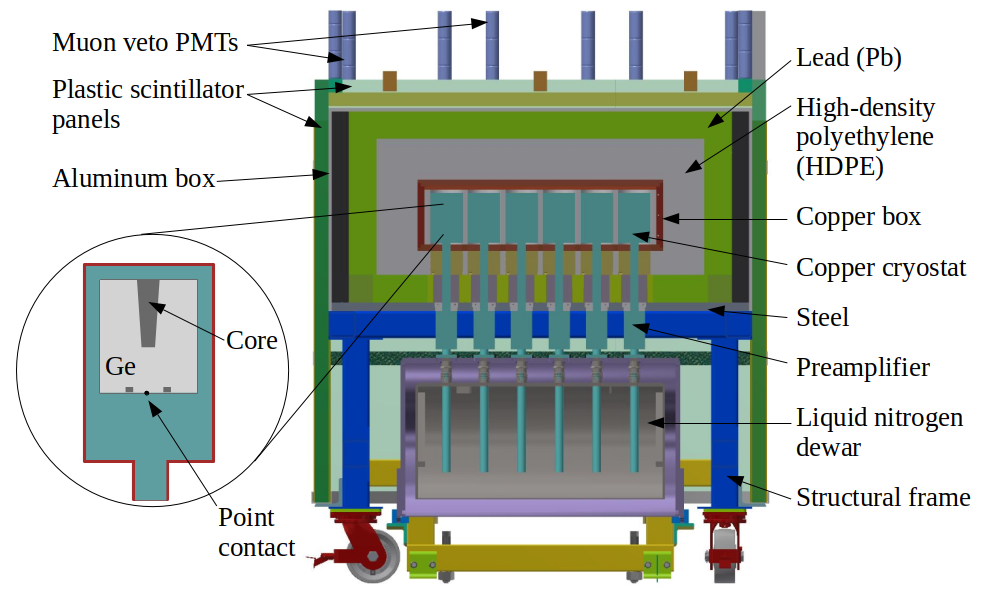}
\caption{\label{fig1:geminisetup} Schematic view of the Ge-Mini detector system: the detectors are enclosed by a radiation shield surrounded on five sides by plastic scintillator panels read out with PMTs used as a muon veto.  
The inset shows the ICPC design of the Ge crystals. Ge-Mini has a total capacity for 12 ICPC detectors.  
}
\end{figure}

\paragraph*{Experiment.-}
The Ge-Mini detectors are p-type crystals with a small-diameter (10\,mm) boron-implanted p+ contacts on the bottom, and lithium-diffused n+ contacts surrounding the majority of the crystal. The central bore-holes (cores) shape internal electric fields to aid depletion (see inset in Figure \ref{fig1:geminisetup}).
This configuration combined with a Transistor-Reset Preamplifier (TRP) sans feedback resistor results in low electronic noise, as confirmed by measurements of the noise FWHM ranging between detectors from 90-140\,eV$_{\textrm{ee}}$\footnote{Electron-equivalent energy measured as ionization}.
This low noise and the excellent intrinsic resolution of HPGe detectors yield a total energy resolution of 220-280\,eV$_{\textrm{ee}}$ FWHM at 10.37\,keV (decay of $^{68}$Ge/$^{71}$Ge). 

The Ge-Mini setup is located at (19.2$\pm$0.1)\,m from the SNS target at a scattering angle of $\sim$90$^{\circ}$ from the beam axis, at the same location as COHERENT's previously deployed CsI detector (see Fig.~2 in \cite{akimov2017observation} for reference). Figure \ref{fig1:geminisetup} shows a schematic overview of the setup; a layered radiological shield with an active muon veto system suppresses backgrounds from natural radioactivity and from the SNS environment, muon-induced secondaries, and beam-related neutrons.  

The data acquisition system (DAQ) consists of two Struck SIS3316\footnote{ 
http://www.struck.de/sis3316.html} 16-bit 125\,MHz FADC cards and we employ the Rdigdaq software\footnote{developed by James Matta for the ORNL Radiation Detection and Imaging Group for the DOE NNSA Office of Defense Nuclear Nonproliferation R\&D.}.

\paragraph*{Data taking.-} 
The data for this analysis were acquired during the beam operating period between June 21, 2023 -- August 15, 2023 throughout which a stable beam energy of 1,050\,MeV was maintained. COHERENT Monte Carlo simulation studies estimate a production rate of in total (0.288$\pm$0.029) neutrinos per proton on target (POT) at this energy \cite{akimov2022simulating}.
During the campaign the beam power was increased from 1.5\,MW to 1.7\,MW.  

During normal SNS operation, germanium waveforms (trace length 176\,$\mu$s) are acquired with two external triggers per beam spill (referred to as externally triggered): 
one in coincidence with the beam spill (on-beam) and one delayed by 1.67\,ms for quantifying steady-state backgrounds (off-beam). This separation allows for a natural blinding scheme with blinded on-beam waveforms, while off-beam waveforms are open for development of the analysis. 
In each waveform the time period ranging from 4\,$\mu$s before the trigger to 36\,$\mu$s after is used for the analysis, ensuring optimal energy resolution avoiding waveform edge effects.

During weekly routine beam-maintenance periods, self-triggered calibration data are collected (referred to as internally triggered) for a high-statistics measurement of the background, including spectral lines used for energy calibration. Five of the seven detectors installed in the shield during the campaign were deemed stable enough for the analysis; the  remaining two detectors exhibited a leakage current too high to be reliably analyzed. The surface layer of the diodes is not active, reducing the active mass by $\sim$4\% to (10.66$\pm$0.09)\,kg. This value is derived from the manufacturer supplied information. The uncertainty is a subdominant contribution (see Table \ref{tab1:uncertainties}), and will be measured precisely for future analyses.

\paragraph*{Data analysis.-}

\begin{figure}
\includegraphics[width=0.45\textwidth]{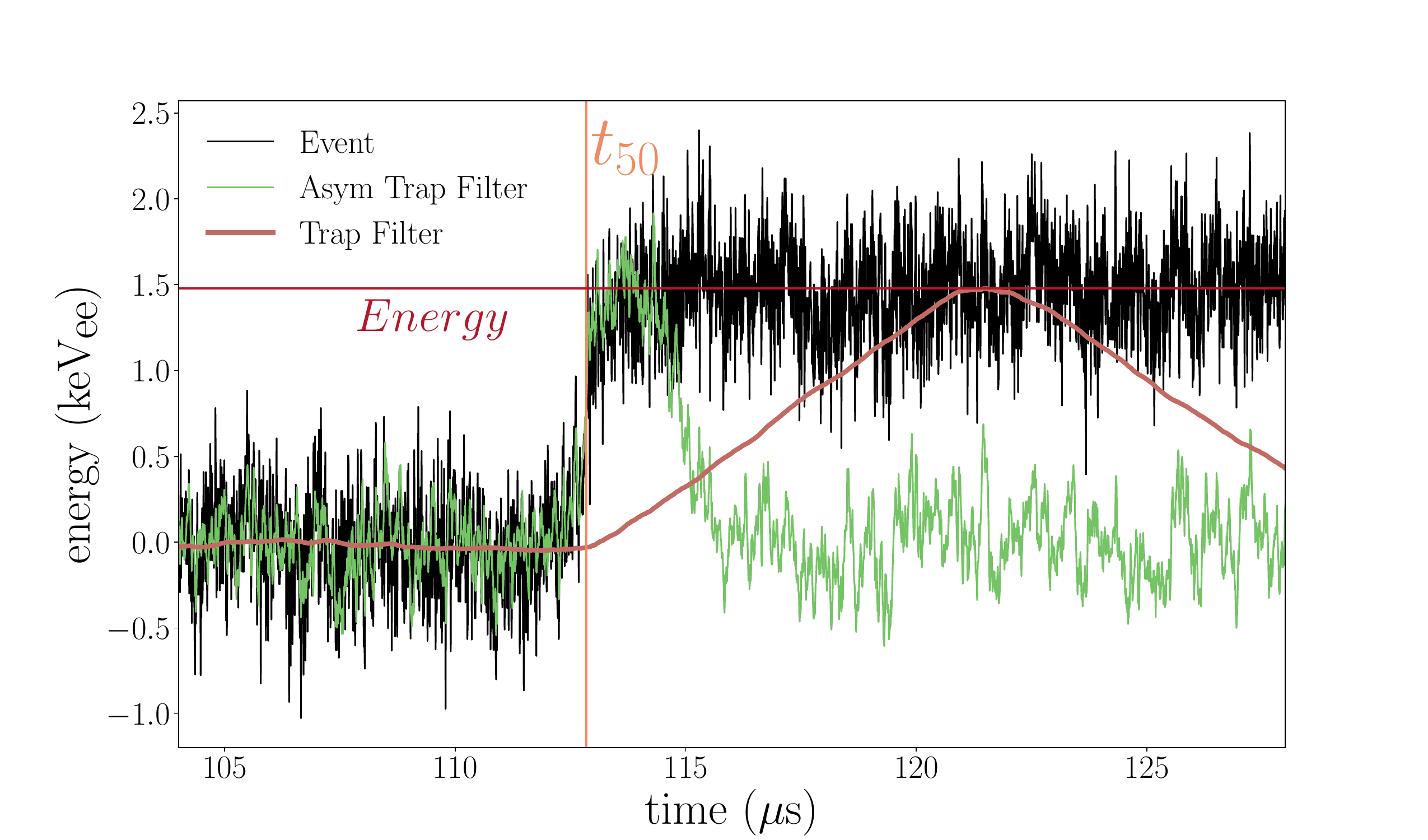}
\caption{\label{fig2:pulse} Example pulse at the analysis threshold of 1.5\,keV$_{\textrm{ee}}$ (black) compared to the pulse after applying the symmetric trapezoidal filter (red) to derive the energy value (red horizontal line) and to the pulse after applying the asymmetric trapezoidal filter (green) to extract t$_{50}$, the time of the arrival of the charges at the readout (orange vertical line). 
}
\end{figure}

\begin{figure}
\includegraphics[width=0.45\textwidth]{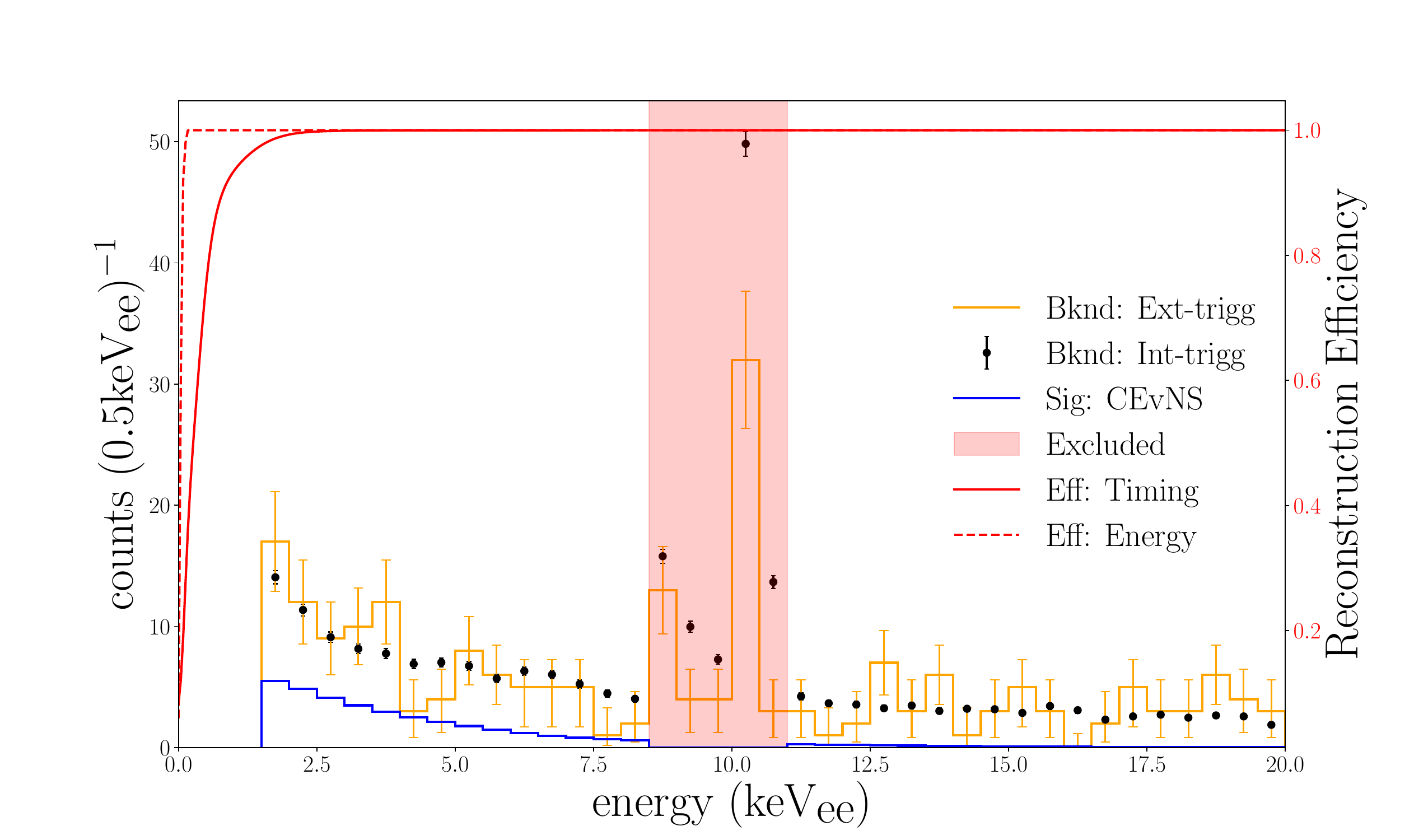}
\caption{\label{fig2:backgroundspectrum}
The steady-state background spectral shape (40\,$\mu$s window) agrees for externally-triggered and rescaled internally-triggered background (used in likelihood fit) within statistics. The red curves indicate the reconstruction efficiencies of energy and timing. The blue histogram illustrates the expected CEvNS signal count rate. 
}
\end{figure}

The event reconstruction and noise cuts are optimized and studied individually for each detector.
Two separate trapezoidal filters \cite{jordanov1994digital} are applied to each germanium waveform to extract energy and arrival time of the induced charge at the readout electrode; an example signal is displayed in Fig.~\ref{fig2:pulse}.
The event energy is proportional to the step increase over the linearly rising raw-waveform baseline; the trapezoidal filter removes the slope of the baseline and averages over noise fluctuations. 
The timing is reconstructed through a method similar to that described in \cite{arnquist2022charge} using an asymmetric trapezoidal filter. From comparing the reconstructed energies of simulated pulses to their actual energies, we found that the energy is reconstructed correctly and follows a linear scale down to 0.2\,keV$_{\textrm{ee}}$. At energies above 2.0\,keV$_{\textrm{ee}}$, the width of the reconstructed pulse onset distribution corresponds to $\pm$230\,ns (3\,$\sigma$), independent of the energy. 
Below 2.0\,keV$_{\textrm{ee}}$, electronic noise begins to impact the timing reconstruction of the waveforms significantly. The resulting efficiency loss at low energies is depicted in Fig.~\ref{fig2:backgroundspectrum}.
Therefore, in combination with the noise performance of the detectors, we set the lower energy threshold of 1.5\,keV$_{\textrm{ee}}$ for our region-of-interest (ROI) in this analysis, resulting in an efficiency loss due to the timing reconstruction of less than 1\%.

A series of analysis cuts are applied to remove spurious events. Within the TRP, charges are accumulated until the maximum of the dynamic range is reached, at which time a reset is automatically triggered, draining charge from the integrator circuit. 
We remove all events in a 200\,$\mu$s window following a reset.
Resets are largely driven by the steady-state leakage current in each detector (time-averaged TRP reset rate of 20-200\,Hz during normal operation, resulting in 0.4-4\% loss of live-time). As the detector leakage current is correlated with electronic noise, time periods with excessive reset rates are removed. Periods of increased noise are also identified by measuring the root-mean-square of the baseline. 
Moreover, the daily liquid nitrogen (LN) fills induce low-frequency noise in sensitive detector electronics. 
LN fills were logged, and the approximately hour-long fills were excluded from the analysis. For two of the detectors these periods were extended up to 30\,min after a fill due to waveforms exhibiting low-frequency oscillations. We also excluded three four second long time periods with excessive rate with several hundred events within less than 1\,s, while less than one background event per day is expected.
Waveforms containing non-physical events are identified and removed if the minimum value of the trapezoidal filter used in the energy reconstruction is 50-500\% smaller than expected for waveforms containing physics events.
An event is cut from the analysis if a muon veto trigger was registered within [-16,16]$\mu$s around the Ge waveform beam window. 
The length of the muon veto coincidence window was optimized with the help of MCNP simulations \cite{TechReport_2023_LANL_MNCP} to exclude all prompt muon-induced secondaries.
Finally, beam triggers were retained in the dataset only if the number of beam event triggers was similar to the number of beam spills with protons-on-target to within 10\%.
The combined effect of all cuts results in a detector-dependent efficiency loss on the live-time of 13-23\%.
Additionally, we cut periods with unstable beam operations.
A total combined exposure after all cuts of 10.22\,GWhkg (2.09$\cdot$10$^{23}$\,POT$\cdot$kg) was acquired during Campaign-2. After unblinding the exposure calculation needed to be corrected due to software bugs, no changes were made to cuts or waveform reconstruction parameters. The resulting background energy spectrum in the CEvNS signal ROI is shown in Fig.~\ref{fig2:backgroundspectrum}.
Due to the small overburden of the experiment, the background before cuts is dominated by muon-induced events. After applying the muon veto, the total background is reduced by more than 80\%.
The observed peaks in Fig.~\ref{fig2:backgroundspectrum} correspond to the K-shell electron-capture decay of the isotopes $^{68}$Ge/$^{71}$Ge (10.37\,keV) and $^{65}$Zn (8.98\,keV) from cosmogenic activation and thermal neutron capture.

Figure \ref{fig2:backgroundspectrum} also confirms that the shape of the internally- and externally-triggered off-beam steady-state background spectra agree outside of the cosmogenic-induced background lines, which are excluded from the normalization and from the ROI of the CEvNS analysis. The upper end of the ROI is restricted to 20\,keV$_{\textrm{ee}}$ (loss of 0.4\% of expected CEvNS interactions). No loss of efficiency in the self-triggered data due to the threshold in the trigger algorithm is observed above the analysis threshold of 1.5\,keV$_{\textrm{ee}}$ for three of the five detectors. The spectra of these three detectors were combined to create the spectrum in Fig.~\ref{fig2:backgroundspectrum}. 
Due to the agreement in shape, confirmed by a Kolmogorov–Smirnov test, \cite{murphy1968handbook}, we employ the internally-triggered spectrum as the background Probability Density Function (PDF) for the likelihood analysis due to its improved statistical precision compared to the externally-triggered data. 
We assume that the background events are distributed uniformly in time, confirmed by the histograms of the time differences between events.

For the internally-triggered data, in the ROI we measure a steady-state background rate of (102$\pm$1)\,counts/d/kg (averaged over all detectors).
This is significantly reduced by the correlation with the SNS beam. In the externally-triggered data only (0.08$\pm$0.02)\,counts/d/kg are observed for a beam window of 10\,$\mu$s (optimized for a counting analysis). This results in an expected CEvNS signal-to-background ratio of $\sim1$.
The beam-related neutron background 
is estimated from the measured neutron spectrum in Neutrino Alley (appendix of \cite{akimov2017observation}) propagated through the shield geometry in a MCNP-based simulation. 
For the full exposure, (0.67$\pm$0.34)\,counts are expected (less than 2\% of the expected CEvNS signal). 
Therefore, a neutron PDF was not included in the on-beam data likelihood analysis. 
The low number of counts is consistent with the CsI data collected at the same location within Neutrino Alley \cite{akimov2017observation,akimov2022measurement}.
The total neutrino-induced neutron contribution from the lead shield is expected to be $<$0.002\, counts based on the CsI measurements \cite{akimov2017observation} and the additional neutron-moderating shielding of Ge-Mini.

The energy scale is calibrated using the 10.37\,keV  peak and the 511\,keV positron-annihilation peak from the internally-triggered data. These data establish an uncertainty on the energy scale of 20\,eV$_{\textrm{ee}}$ within the ROI.

\paragraph*{Signal expectation.-} The timing of the germanium detectors must be evaluated precisely to correlate detected events with the production of neutrinos at SNS.
Drift-times in the large diodes vary by several $\mu$s depending on where the charge cloud is produced.
We determined the detector-dependent drift-time distributions by simulating pulse shapes in a grid in each detector using the germanium detector modeling codes fieldgen and siggen \cite{Radford2014}.
We validated these simulations and measured the timing contribution of the electronics through coincidence measurements with a gamma-ray source placed between a bismuth germanate detector and the Ge diodes. 
A combination of simulation and measurement results is used to produce the expected drift-time distribution for uniformly distributed events in each detector; this is convolved with the expected time distributions to generate the signal time distributions.
More than 99\% of the neutrino signals are expected to be read out within 14\,$\mu$s after a beam spill. 


Germanium spectrometers only measure ionization, while nuclear recoil's energy depositions are divided between ionization and phonon production.  
This quenching effect needs to be taken into account when calculating the expected CEvNS spectrum and rate. It is described by the semi-empirical Lindhard theory \cite{Lindhard1963} as confirmed by \cite{bonhomme2022direct} and \cite{li2022measurement}. We set the only free parameter of the Lindhard theory to $k$=0.157, corresponding to the predicted value in \cite{Lindhard1963}. Due to the nuclear recoil quenching, the analysis threshold of 1.5\,keV$_{\textrm{ee}}$ corresponds to a recoil energy of $\sim$6.7\,keV$_{nr}$.
We assume an uncertainty on $k$ of 0.004 as 
quoted for the measurement in \cite{bonhomme2022direct}. The impact on our analysis is negligible compared to the other systematic uncertainties (see Table \ref{tab1:uncertainties}), this energy regime is above the range where deviations between quenching factor measurements persist \cite{bonhomme2022direct, li2022measurement, collar2021germanium} (and result in difficulties in measurement interpretation). Thus there is no systematic uncertainty associated to quenching here.
At higher neutrino energies (several tens of MeV), the shape of the nucleus, described by the nuclear form factor, starts to impact the shape of the recoil spectrum and the flux-weighted signal expectation by reducing it to (68$\pm$1)\% \cite{helm1956inelastic,klein1999exclusive}.

\begin{table}
    \centering
    \begin{tabular}{l|c}
       Uncertainty  &  contribution\\ \hline
       flux & 10\% \\
       distance & 0.5\% \\
       energy calibration  & 1\%\\
       active mass  & 2\%\\
       form factor  & 1\%\\
       quenching Ge & negligible \\ \hline
       all &10.3\%\\
    \end{tabular}
    \caption{Overview of systematic uncertainties on the signal rate. The SNS neutrino flux will be addressed in an upcoming D$_2$O measurement \cite{akimov2021d2o}.}
    \label{tab1:uncertainties}
\end{table}

\paragraph*{Results.-}

\begin{figure}
\includegraphics[width=0.5\textwidth]{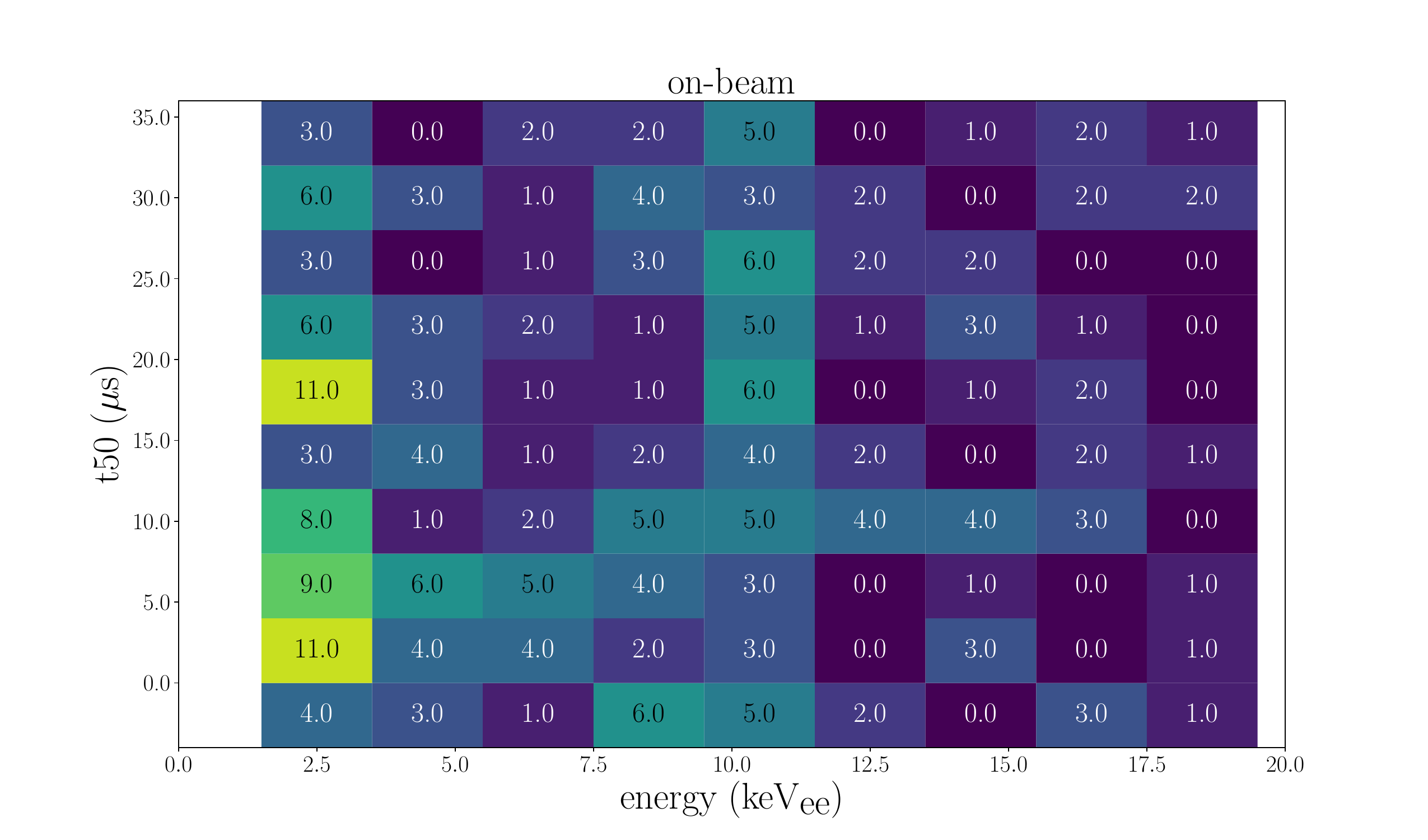}
\includegraphics[width=0.5\textwidth]{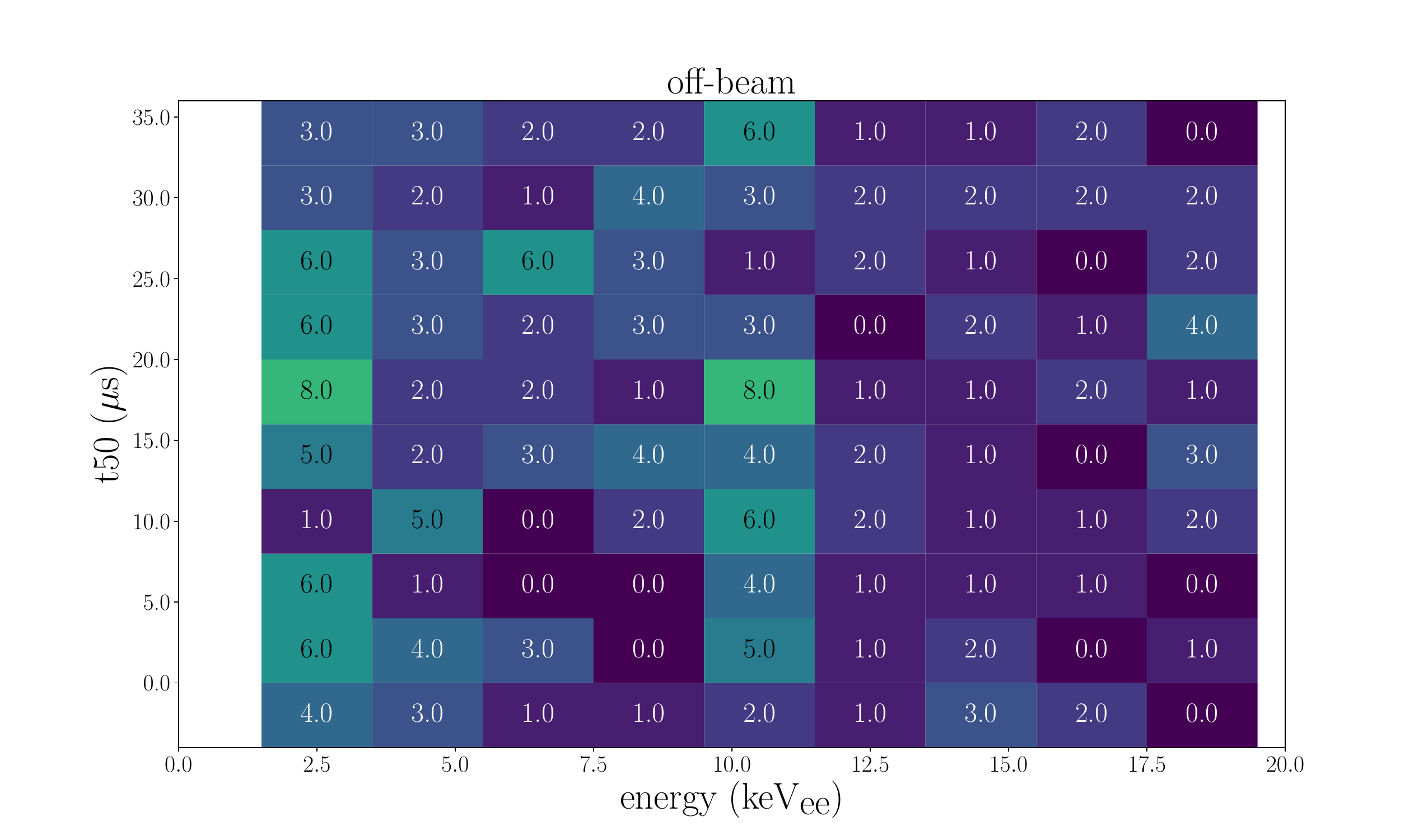}
\caption{\label{fignew}
Top: on-beam data including the CEvNS related excess; bottom: off-beam data flat within statistical fluctuations.
}
\end{figure}

\begin{figure*}
\includegraphics[width=0.9\textwidth]{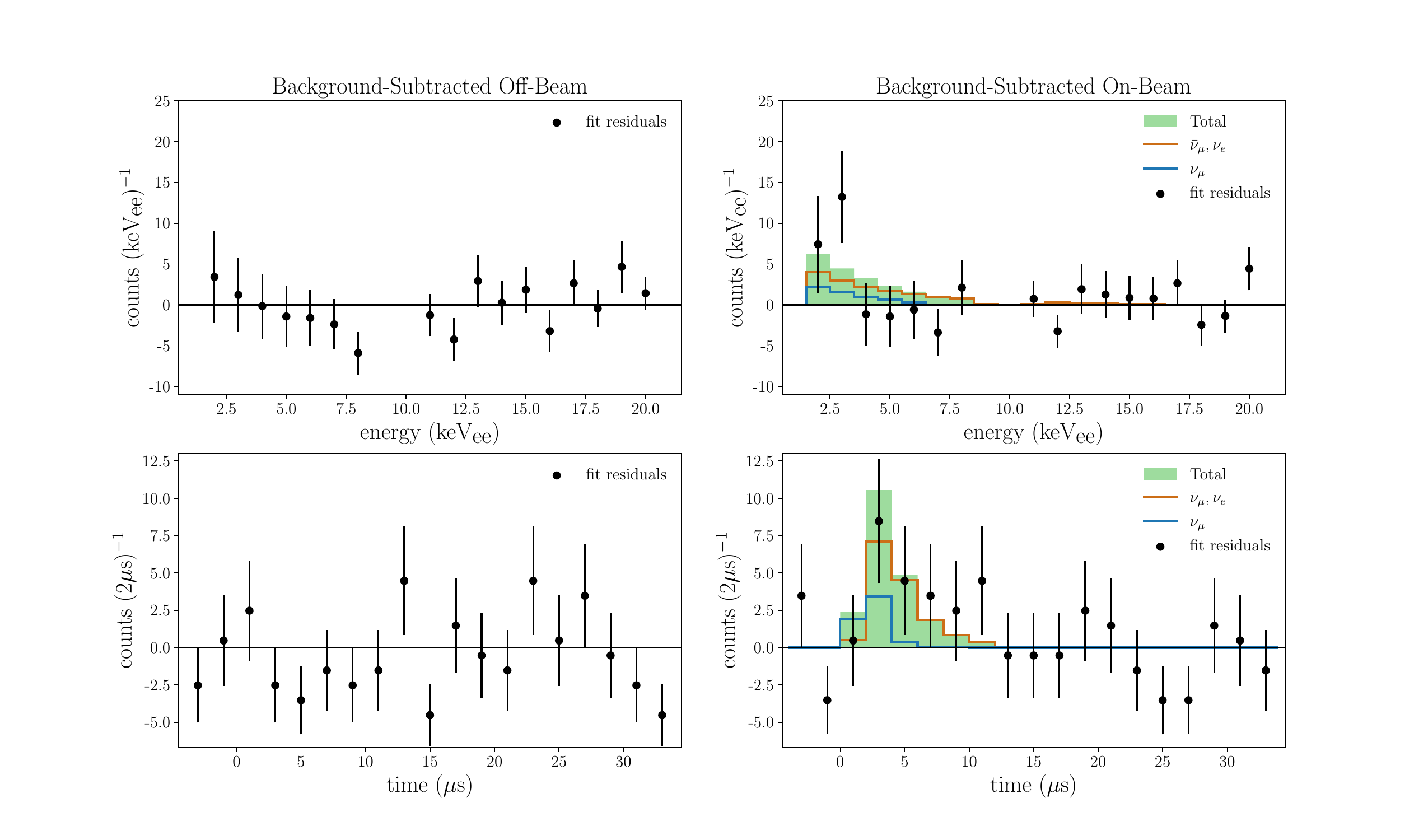}
\caption{\label{fig4:cevnsfitresult} Background-subtracted spectra in time and energy for on-beam and off-beam data. The signal extracted by the likelihood fit is depicted in green; the red and blue lines represent the different neutrino flavors.  
}
\end{figure*}

We performed an unbinned, two-dimensional (energy, time) maximum likelihood fit simultaneously on the off-beam and on-beam data. The only two fit parameters are the numbers of signal and background counts, the latter constrained by the simultaneous fit to the off-beam data. The fit range excludes the cosmogenic lines 
within $[8.5,11.0]$\,keV$_{\textrm{ee}}$.  
We apply the full 40\,$\mu$s acceptance window around the beam trigger to maximally constrain the steady-state backgrounds. Fig.~\ref{fignew} directly depicts the 2D spectra of the on-beam data with the CEvNS related excess and the off-beam data. Figure~\ref{fig4:cevnsfitresult} shows the data after subtraction of the background PDF scaled by its fit value; the presence of a CEvNS signal is statistically supported by the on-beam data.

\begin{figure}
\includegraphics[width=0.48\textwidth]{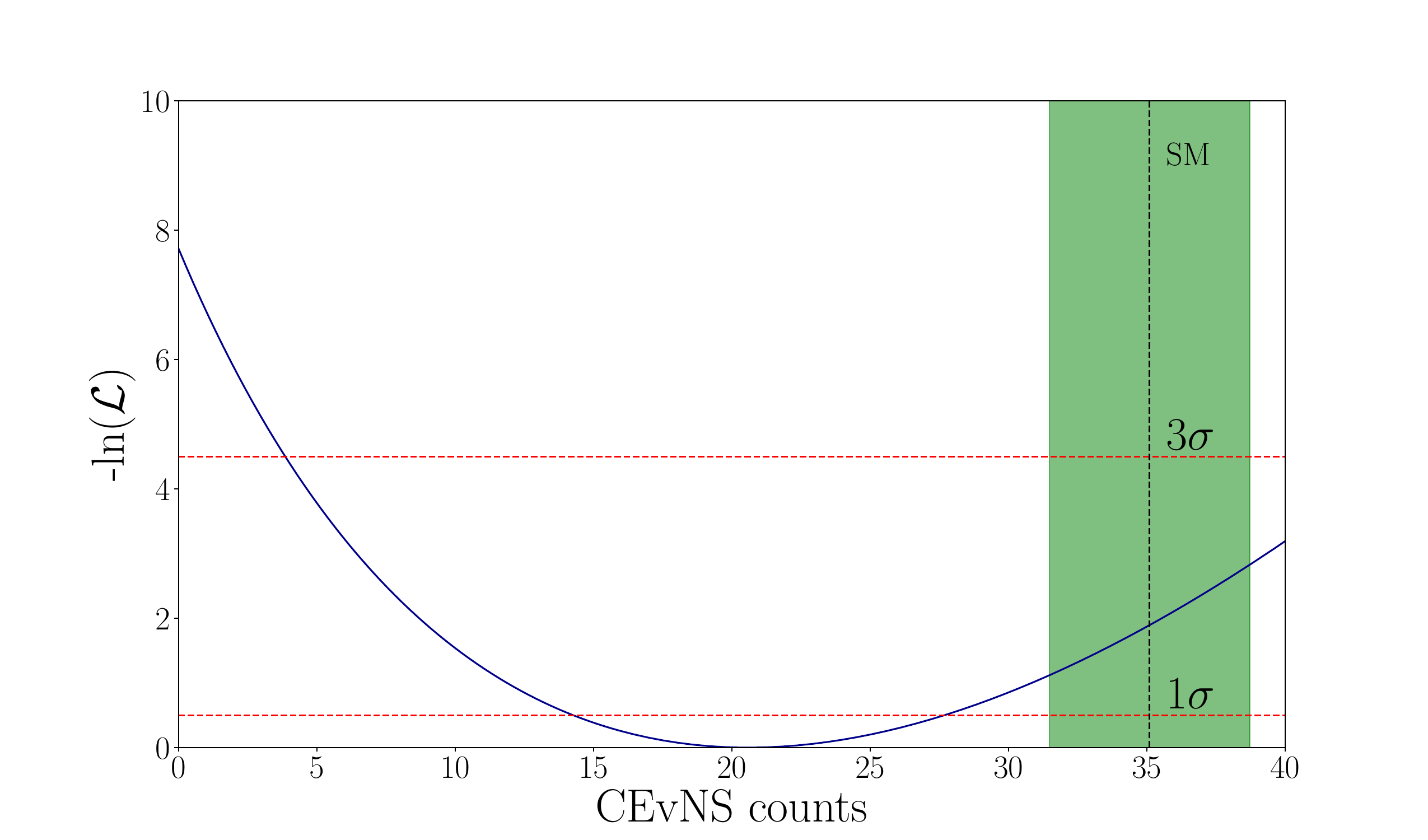}
\caption{\label{fig5:profilelikelihood} The likelihood profile has its minimum at the number of CEvNS counts derived from the fit. 
The black-dotted line represents the SM signal expectation. The green band corresponds to the total systematic uncertainty.}
\end{figure}

Figure~\ref{fig5:profilelikelihood} shows the negative natural logarithm of the likelihood as a function of CEvNS counts. 
In total 20.6$_{-6.3}^{+7.1}$ CEvNS-signal-like events are observed over steady-state background, of which 0.67$\pm$0.34 are estimated to result from beam-related neutrons.
The null hypothesis of zero CEvNS signal is rejected at 3.9\,$\sigma$, derived from a one-sided $\chi^2$-distribution as confirmed by toy MC \cite{feldman1998unified}. The goodness-of-fit (reduced $\chi^{2}$) equals 1.84 ($p$=0.40).
The 2D likelihood analysis is consistent with a counting analysis performed over a 2D range spanning $[1.5,8.5]$\,keV$_{\textrm{ee}}$ and $[0,8]$\,$\mu$s; 21.0$\pm$7.8 signal-like events were found in the data.

The standard model signal prediction for our exposure of 35.1$\pm$3.6 (applied form factor model \cite{klein1999exclusive}), including systematic uncertainties, is shown as green band in Fig.~\ref{fig5:profilelikelihood}.
Our measured 20.6$^{+7.1}_{-6.3}$ counts, including the small beam-related neutron component of less than 2\%, are within 1.95\,$\sigma$ of the standard model prediction.

In this dataset, we are limited by the statistical uncertainty of the measured data, though future operation at the SNS will reduce this significantly. 
Any shape uncertainties on the fitted PDFs are negligible in comparison. The included systematic uncertainties regarding the neutrino flux, the detector properties and performance as well as the theoretical knowledge of neutrino interactions on Ge amount to 10.3\% as summarized in Table \ref{tab1:uncertainties}. The dominant contribution is the uncertainty in the neutrino production rate at the SNS. There are efforts within the COHERENT collaboration to significantly reduce this uncertainty by independently measuring the neutrino flux with charged-current deuteron scattering in a D$_2$O Cherenkov detector \cite{akimov2021d2o}.

\paragraph*{Summary and outlook.-}
With a total exposure of 10.22\,GWhkg and an analysis threshold of 1.5\,keV$_{\textrm{ee}}$ we report the first-ever detection of CEvNS on germanium nuclei with a significance of 3.9\,$\sigma$. 

A further careful study of the noise-dominated pulse shapes may make it possible to extend the analysis to lower, still-blinded energy regions with strongly increasing signal expectation. 
Moreover, a significant number of slow pulses are present in the dataset from interactions in the surface layers of the diode. 
A pulse-shape discrimination algorithm (similar to e.g., \cite{cooper2011pulse}, \cite{agostini2022pulse} and \cite{bonet2024pulse}) may be used to remove these events.
Finally, an increase in exposure from additional measurement time and detector mass will significantly reduce the statistical uncertainty of the result. 
The SNS Proton Power Upgrade, completed in 2024 \cite{akimov2022simulating}, will gradually increase the beam power to 2.0\,MW, further enhancing the neutrino flux. Improved exposure and reduced systematic uncertainties will enable precision tests of the SM with neutrino-induced recoils and significantly improve limits on beyond-the-SM physics.
\paragraph*{Acknowledgements.-}
The COHERENT Collaboration acknowledges the generous resources provided by the ORNL Spallation Neutron Source, a DOE Office of Science User Facility. Laboratory Directed Research and Development funds from ORNL also supported this project. We acknowledge support from U.S. Department of Energy Office of Science and the National Science Foundation.  
The Ge-Mini array detectors and hardware were acquired through National Science Foundation Major Research Instrumentation Award Number 1920001.
We also acknowledge support from the Alfred P. Sloan Foundation, the Consortium for Nonproliferation Enabling Capabilities, and the Korea National Research Foundation (No. NRF 2022R1A3B1078756). This research used the Oak Ridge Leadership Computing Facility, which is a DOE Office of Science User Facility. We also acknowledge support from Ministry of Science and Higher Education of the Russian Federation, Project “New Phenomena in Particle Physics and the Early Universe” FSWU-2023-0073. We thank Mirion Technologies, Meriden CT, USA for guidance and support with the detectors.

\nocite{*}
\bibliography{apssamp}

\end{document}